\documentclass[aps,prb,twocolumn,superscriptaddress,showpacs]{revtex4}

\usepackage{graphicx}   % need for figures
\usepackage{longtable}

\begin{document}

\preprint{accepted to PRB}

\title{Laser-Shock Compression and Hugoniot Measurements of Liquid
  Hydrogen to 55 GPa}


\author{T. Sano}
%\thanks{}
\affiliation{Institute of Laser Engineering, Osaka University, Suita,
  Osaka 565-0871, Japan} 

\author{N. Ozaki}
\affiliation{Graduate School of Engineering, Osaka University,
Suita, Osaka 565-0871, Japan}
\author{T. Sakaiya}
\affiliation{Department of Earth and Space Science, Osaka University,
  Toyonaka, Osaka 560-0043, Japan} 
\author{K. Shigemori}
\affiliation{Institute of Laser Engineering, Osaka University, Suita,
  Osaka 565-0871, Japan} 
\author{M. Ikoma}
\affiliation{Department of Earth and Planetary Sciences, Tokyo
  Institute of Technology, Ookayama, Meguro, Tokyo 152-8551, Japan}
\author{T. Kimura}
\affiliation{Graduate School of Engineering, Osaka University, Suita,
  Osaka 565-0871, Japan} 
\author{K. Miyanishi}
\affiliation{Graduate School of Engineering, Osaka University, Suita,
  Osaka 565-0871, Japan} 
\author{T. Endo}
\affiliation{Graduate School of Engineering, Osaka University, Suita,
  Osaka 565-0871, Japan} 
\author{A. Shiroshita}
\affiliation{Institute of Laser Engineering, Osaka University, Suita,
  Osaka 565-0871, Japan} 
\author{H. Takahashi}
\affiliation{Department of Earth and Space Science, Osaka University,
  Toyonaka, Osaka 560-0043, Japan} 
\author{T. Jitsui}
\affiliation{Graduate School of Engineering, Osaka University, Suita,
  Osaka 565-0871, Japan} 
\author{Y. Hori}
\affiliation{Department of Earth and Planetary Sciences, Tokyo
  Institute of Technology, Ookayama, Meguro, Tokyo 152-8551, Japan}
\author{Y. Hironaka}
\affiliation{Institute of Laser Engineering, Osaka University, Suita,
  Osaka 565-0871, Japan} 
\author{A. Iwamoto}
\affiliation{National Institute of Fusion Science, Toki, Gifu
  509-5292, Japan} 
\author{T. Kadono}
\affiliation{Institute of Laser Engineering, Osaka University, Suita,
  Osaka 565-0871, Japan} 
\author{M. Nakai}
\affiliation{Institute of Laser Engineering, Osaka University, Suita,
  Osaka 565-0871, Japan} 
\author{T. Okuchi}
\affiliation{Institute for Study of the Earth's Interior, Okayama
  University, Misasa, Tottori 682-0193, Japan}
\author{K. Otani}
\affiliation{Institute of Laser Engineering, Osaka University, Suita,
  Osaka 565-0871, Japan} 
\affiliation{Advanced Research Center for Beam Science, Institute for
  Chemical Research, Kyoto University, Uji, Kyoto 611-0011, Japan}
\author{K. Shimizu}
\affiliation{KYOKUGEN, Center for Quantum Science and Technology under
  Extreme Conditions, Osaka University, Toyonaka, Osaka 560-8531,
  Japan} 
\author{T. Kondo}
\affiliation{Department of Earth and Space Science, Osaka University,
  Toyonaka, Osaka 560-0043, Japan} 
\author{R. Kodama}
\affiliation{Graduate School of Engineering, Osaka University, Suita,
  Osaka 565-0871, Japan} 
\author{K. Mima}
\affiliation{Institute of Laser Engineering, Osaka University, Suita,
  Osaka 565-0871, Japan} 


\date{\today}

\begin{abstract}
The principal Hugoniot for liquid hydrogen was obtained up to 55 GPa
under laser-driven shock loading. 
Pressure and density of compressed hydrogen were determined by
impedance-matching to a quartz standard. 
The shock temperature was independently measured from the brightness
of the shock front.  
% Nov 21
% Hugoniot data of hydrogen are in good agreement with theoretical
% predictions and quantitatively different from the deuterium results. 
Hugoniot data of hydrogen provide a good benchmark to modern theories
of condensed matter.
% Dec 5
% and quantitatively different from the deuterium results.  
The initial number density of liquid hydrogen is lower than that for liquid
deuterium, and this results in shock compressed hydrogen having a
higher compression and higher temperature than deuterium at the same
shock pressure.
% Nov 21
% The initial density effects have been detected, which give higher
% compression and higher Hugoniot temperature of hydrogen compared to 
% those of deuterium at the same pressure. 
\end{abstract}

%\pacs{62.50.-p, 64.30.-t, 52.72.+v, 71.30.+h}
\pacs{62.50.-p  51.30.+i  52.72.+v  64.30.-t}
%\keywords{}

\maketitle

%\section{}
%\subsection{}
%\subsubsection{}

\section{INTRODUCTION}

The properties of hydrogen at high pressure and high density are of
great scientific interest. 
The equation of state (EOS) of hydrogen at these conditions is
essential for modeling of the interior structure of gas giant planets
%\cite{saumon04}. 
\cite{saumon04,nettelmann08,militzer08}. 
The large diversity in the estimation of Jupiter's core mass is
resulted from the uncertainty in the EOS data especially in the region
around the insulator-to-metal transition. 
The EOS of hydrogen-isotopes has important practical applications for
inertial confinement fusion \cite{lindl95}, and metallic hydrogen is
suggested as a prospective candidate of high-temperature
superconductor \cite{ashcroft68}. 
%\cite{ashcroft68,richardson97}. 
Chemical free-energy models \cite{saumon95,ross98,kerley03} and $ab$
$initio$ simulations \cite{collins01prb,holst08} have been used to
predict the properties of warm dense hydrogen, but the results vary
widely and have not converged yet. 
Therefore accurate experimental data for the hydrogen EOS are required
for evaluation of the theoretical models and for further understanding
of the fundamental nature of hydrogen. 

% It is more difficult to generate high pressures in H$_2$ than in
% D$_2$ because of its lower shock impedance.  
It is more difficult to generate high pressures in hydrogen than in
deuterium because of its lower shock impedance.  
For this reason, most of the recent experimental measurements by shock
compression have focused on the heavier isotope
%\cite{mostovych00,boehly04,knudson04,hicks09}. 
\cite{collins98,mostovych00,boehly04,knudson04,boriskov05,hicks09}. 
However, it should be noted that the Hugoniots for the two isotopes do
not scale in density.
Owing to the difference in zero-point energy, the mole volume of
liquid hydrogen is larger than deuterium \cite{wiberg55,kerley03}.
%, and the dissociation energy is lower
% As a result, H$_2$ is expected to have a higher Hugoniot temperature
% and higher compression than D$_2$ at the same pressure.
As a result, hydrogen is expected to have higher compression and
higher Hugoniot temperature than deuterium at the same pressure.  

There is a large gap in the experimental achievement of shock
compression between liquid hydrogen and deuterium. 
The principal Hugoniot for liquid deuterium was measured up to 220 GPa
using laser-driven shock waves \cite{hicks09}. 
For the case of liquid hydrogen, the Hugoniot was studied
experimentally only to 10 GPa by a gas gun and explosive method more
than two decades ago \cite{dick80,nellis83}.  
The metallization of hydrogen on the Hugoniot is expected to occur at
much higher pressure.
% Nov 21
% It was reported recently that solid hydrogen was compressed up to 65
% GPa by using spherical explosive charges \cite{trunin05}.
%%To compensate for the lack of hydrogen data, we carried out
%%laser-shock experiments of liquid hydrogen. 
In this work, we carried out laser-shock experiments of liquid
hydrogen to pressures exceeding 10 GPa in order to make a
quantitative comparison of the hydrogen Hugoniot around the metal
transition with the deuterium data.  
%% In this letter, we report laser-shock experiments of liquid hydrogen.
%% The primary aim of this work is to reach pressures of liquid hydrogen
%% substantially exceeding those achieved so far and to make a
%% quantitative comparison of the Hugoniot data between H$_2$ and D$_2$.

\section{EXPERIMENTS}

The experiments were performed on the GEKKO HIPER laser facility at
the Institute of Laser Engineering, Osaka University. 
The laser is a neodymium-doped glass system operating at the third
harmonics wavelength of 351 nm. 
Laser energies between 0.8 and 1.4 kJ were delivered 
to generate the shock pressures
using a nominally square pulse of 2.5 ns in duration.
The laser focal spot of 600 $\mu$m in diameter was smoothed using
Kinoform phase plates. 
This resulted in the effective laser intensities between 4 and $8
\times 10^{13}$ W/cm$^2$.

Figure 1 shows the experimental setup and target arrangement.  
The cryogenic hydrogen target consisted of two z-cut
$\alpha$-quartz plates, by which the hydrogen layer was sandwiched.
The thickness of the quartz was 50 $\mu$m for both of them.
These quartz plates were glued to kovar flanges attached to a copper
cell filled with liquid molecular hydrogen at 15 K.  
The gap between the two quartz windows was 30-250 $\mu$m thick. 
%The front-side of the quartz, illuminated by the laser drive, 
%The laser-side quartz was deposited by 40 $\mu$m aluminum to reduce
%the x-ray radiation from ablation plasma.
The laser-side quartz was deposited by 40 $\mu$m aluminum as a pusher.
The aluminum layer has another important role to shield the quartz and
hydrogen against the x-ray radiation from ablation plasma. 
The free surface of the rear-side quartz had an anti-reflective
coating to minimize spurious reflections.
%%from the surface.
%coating to minimize back reflections.  
In this target, the quartz was designed to be used not only as a window
material but also as a reference standard for the impedance-matching
measurement. 
%Thickness of the hydrogen layer is measured by interferometry using
%white light at room temperature \cite{letoullec89}.
The initial density of quartz and liquid hydrogen were 2.65 and 0.0760
g/cm$^3$. 
At the probe laser wavelength 532 nm, the index of refraction of
quartz and hydrogen were 1.547 and 1.121 \cite{souers86}. 
Small change in the optical properties of quartz at cryogenic
temperature is ignorable \cite{hicks09}.
%temperature can be ignored in this work \cite{hicks09}.

\section{RESULTS: SHOCK VELOCITIES}

Shock velocities in the laser-side quartz $U_{s\rm Q}$ and hydrogen
$U_{s\rm H}$ were measured using a line-imaging velocity
interferometer system for any reflector (VISAR)
%\cite{celliers04}.  
\cite{barker72,celliers04}.  
At the high pressure involved in these experiments, shock waves in the
initially transparent quartz and hydrogen are reflecting. 
Then the VISAR can provide a direct time-resolved measurement of shock
velocities in these media.
The probe light for VISAR 
was a Q-switched yttrium-aluminum-garnet (YAG) laser operating at a
wavelength of 532 nm.   
Two VISARs were run concurrently on each shot to resolve the 2$\pi$
phase-shift ambiguities at shock breakout.
The velocity sensitivities were 4.14 and 14.53 km/s/fringe for quartz,
and 5.71 and 20.05 km/s/fringe for liquid hydrogen.
Post processing of the VISAR images using Fourier transform methods
determines the fringe position to $\sim 5$\% of a fringe.
The resulting velocities were measured to $\sim 1$\% precision 
since shock velocities were high enough to cause multiple fringe shifts.  

A sample VISAR trace is shown in Fig. 2(a) and the accompanying
profile of shock velocities in Fig. 2(b). 
%% The shock front was observed first entering the quartz, and 
%% then breaking out of the quartz into liquid hydrogen. 
The time-resolved VISAR measurements allowed velocities to be tracked
continuously during transit through both the standard and sample.
The transit distance in the quartz determined from the time-integrated
VISAR velocity agreed consistently with the thickness measured by
interferometry before the shock experiment.
Shock velocities immediately before and after 
%the shock crossed 
the quartz-hydrogen interface were adopted for the impedance-matching
analysis.
To precisely determine the velocity at the breakout, a linear fit was
taken of the velocities over 0.5 ns and extrapolated to forward or
backward the impedance-matching time $t = 0$.
The velocities were calculated by averaging the shock profile
over 0.1 ns, which is equivalent to the temporal resolution of the
streak camera. 

As seen from Fig. 2(b), the shock velocity in quartz decreased with
time more than 15\% within $\sim 2$ ns.
Decaying shock is a common feature of laser-driven shock waves. 
This means that the estimation of shock velocity by transit time
requires corrections taking account of the deceleration rate, and thus
it could bring a large uncertainty in $U_s$.  
Therefore transparent standards have a great advantage for laser-shock 
experiments, and actually provide a significant improvement in the
accuracy of the evaluation of the pressure $P$ and density $\rho$
along the Hugoniot \cite{hicks08,ozaki09}. 
Shock planarity was observed to extend over 400 $\mu$m which ensures
more than 5 fringes to be available for the Fourier analysis.
In this spatial range, the systematic variation in the shock arrival
time was less than about 0.1 ns.
Preheat in quartz and hydrogen can be neglected because preheat in the
laser-side quartz was estimated to be less than 0.04 eV and no motion
of the quartz-hydrogen interface was observed before shock breakout in
1D radiation hydrodynamic simulations with the MULTI code
\cite{ramis88}. 

Figure 3 shows a impedance-match diagram in the $P$-$U_p$ plane, where
$U_p$ is the particle velocity. 
The quartz Hugoniot was previously established by the OMEGA-laser
experiments to have a linear 
$U_{s}$-$U_{p}$ relation given by $U_{s{\rm Q}} = a_0 + a_1 ( U_{p
{\rm Q}} - \beta)$, where $a_0 = 20.57 \pm 0.15$, $a_1 = 1.291 \pm
0.036$, and $\beta = 12.74$ \cite{hicks05,hicks08}.  
This formula is valid at the pressure over 200 GPa and in good
agreement with the Kerley-7360 EOS model \cite{kerley99}.
Impedance-matching occurs at the intersection of the hydrogen
Rayleigh line and the release isentrope of quartz.
Because of the low shock impedance of hydrogen, quartz release curves
from several hundreds of GPa down to more than one order of magnitude
are needed. 
%from several hundreds of GPa down to very low pressure are needed.
% Note that a phase transition of quartz takes place below 20 GPa
% \cite{wackerle62}.
% Nov 25
% In this work, the quartz release is approximated by a reflection of
% the principal Hugoniot in the $P$-$U_p$ plane \cite{enig63}.
In this work, the quartz isentrope was calculated from the tabulated
Kerley model. 

Pre- and post-shock conditions are related by the Rankine-Hugoniot
conservation equations:
\begin{equation}
\rho_0 U_s = \rho \left( U_s - U_p \right) \;,
\label{eqn:1}
\end{equation}
\begin{equation}
P - P_0 = \rho_0 U_s U_p \;,
\label{eqn:2}
\end{equation}
\begin{equation}
e - e_0 = \frac12 \left( P + P_0 \right) \left( \frac1{\rho_0} -
\frac1{\rho} \right) \;,
\end{equation}
where $e$ is the specific internal energy and subscript 0 denotes
initial conditions.
The pressure and density of shocked hydrogen are derived from
measured $U_s$ and $U_p$ using Eqs. (\ref{eqn:1}) and (\ref{eqn:2}). 
The results for the hydrogen Hugoniot are listed in Table I.
Random errors shown in the table come from measurement uncertainties
in $U_s$.
Random error propagation in the impedance-matching procedure is
illustrated by Fig. 3.  
Systematic errors arising from uncertainties in the principal Hugoniot
of quartz are also considered.
Figure 4 depicts the principal Hugoniot of liquid hydrogen in the
$P$-$\rho$ plane, where $\rho$ is normalized by the initial hydrogen
density $\rho_0$. 
%The error bars correspond to the quadrature sum of random and
%systematic errors. 
These results are shown along with the past experimental data 
\cite{dick80,nellis83} and theoretical models calculated for hydrogen
with the initial density $\rho_0 = 0.071$ g/cm$^3$
\cite{kerley03,holst08,ross09}.

The pressure range of shocked hydrogen obtained in this work was 25-55
GPa. 
The highest pressure is more than 5 times higher than the previous data. 
The compression $\rho/\rho_0$ of hydrogen ranges from 3.8 to
5.6, which is mostly comparable to the models but slightly softer at
30-50 GPa.
%% Note that the model predictions have little difference at this
%% pressure. 
Note that the slope change near 15 GPa in the Hugoniot curve predicted
by the quantum molecular dynamics simulations is related to the
dissociation of hydrogen molecules \cite{bonev04a}. 
The compression of deuterium, on the other hand, 
has been measured as 3.3-4.4 at the same 
pressure range \cite{knudson04,hicks09,holmes95}, so that hydrogen
exhibits systematically higher compression compared to deuterium (see
Fig. 4). 
This trend is qualitatively consistent with theoretical predictions
\cite{kerley03} and can be seen in the earlier data below 10 GPa
\cite{nellis83}. 

The uncertainties in the Hugoniot pressure and density are caused
mostly by the random errors associated with velocity measurements,
$\delta U_s$ and $\delta U_p$, which are evaluated by
\begin{equation}
\frac{\delta P}{P} = \sqrt{ 
\left( \frac{\delta {U_s}}{U_s} \right)^2 +
\left( \frac{\delta {U_p}}{U_p} \right)^2} \;,
%\label{eqn:err}
\end{equation}
\begin{equation}
\frac{\delta {\rho}}{\rho} = 
%\frac{U_p}{U_s - U_p} 
\left( \frac{\rho}{\rho_0} - 1 \right)
\sqrt{ \left( \frac{\delta {U_s}}{U_s} \right)^2 +
\left( \frac{\delta {U_p}}{U_p} \right)^2} \;.
\label{eqn:err}
\end{equation}
As seen from Fig. 4, the uncertainties in density are much larger than
those in pressure. 
Since hydrogen is highly compressible at this pressure range, the
factor in Eq. (\ref{eqn:err}) becomes considerably large, $\rho /
\rho_0 -1 \approx$ 4.
Then the errors in density are enhanced up to about 6\%, even when 
the velocity errors are sufficiently small, $\delta U_s/U_s \sim \delta
U_p/U_p \sim 1$\%.  

\section{RESULTS: TEMPERATURE}

Using a streaked optical pyrometer (SOP), the temperature $T$ of
shocked hydrogen was measured simultaneously \cite{ozaki09,miller07}. 
Despite temperature is fundamental to thermodynamics, $T$ is not a part
of the Rankine-Hugoniot relations, and thus must be measured separately
from shock pressure and density.
We extracted the temperature by fitting the absolute spectral radiance
$I_{\lambda}$ to a gray body Plank spectrum,
\begin{equation}
I_{\lambda}(T) = \varepsilon (\lambda) 
\frac{2 \pi h c^2}{ \lambda^5} \frac1{e^{hc/\lambda k T} - 1} \;,
\end{equation}
with $T$ as a fit parameter, where $c$ is the speed of light, $h$ is
Planck's constant, and $k$ is Boltzmann's constant.

The radiance was obtained in a single spectral band centered at 455 nm
with a bandwidth 38 nm.  
The emissivity $\varepsilon$ is related to the reflectivity $R$ of the 
shock front through $\varepsilon (\lambda) = 1 - R(\lambda)$. 
For the reflectivity at the SOP wavelength, we adopted the
reflectivity measured at the VISAR wavelength assuming a weak
dependence of $R$ on the wavelength (see Fig. 7).  
To relate the diagnostic system output to a source radiance, $in$
$situ$ measurements were performed to determine the spectral response
of the system.  
We used calibration signals recorded with a 3000-K tungsten
quartz-halogen reflectorized lamp to find the radiance at our SOP
wavelength.  
The radiance of this lamp was calibrated with a bolometric calibration
traceable to the National Institute of Standards and Technology (NIST).

Absolute reflectivity is determined by comparing the shock
reflectivity to that from the aluminum surface which has a known
value, $86 \pm 6$\%.  
The obtained $R$ of shock front in hydrogen is listed in Table
II.  
The reflectivity just after the shock breakout was calculated by
the same algorithm as the shock velocity $U_s$ using VISAR signals. 
The reflectivity of shocked hydrogen increases from 8\% to 23\% as the
pressure increases from 25 GPa to 41 GPa. 
% Notice that the reflectivity of shocked quartz in these shots was
% consistent with previous laser experiment \cite{hicks06} and quantum
% molecular dynamics simulations \cite{laudernet04}. 
The Fresnel formula for the reflectivity of the shock front is $R = |
(\hat{n}_s - n_0) / (\hat{n}_s + n_0) |^2$, where $\hat{n}_s$ is the
complex refractive index behind the shock front and $n_0$ is the
refractive index in the undisturbed liquid.
At low pressures, the refractive index of liquid hydrogen follows an
empirical form, $n-1 \propto \rho$ \cite{souers86}. 
% empirical Gladstone-Dale form, $n-1 \propto \rho$ \cite{souers86}. 
Assuming the fluid remains mostly in its molecular form under shock
compression, the refractive index at fivefold compression would be $n
\sim 1.6$, and the reflectivity $\sim 4$\%. 
The much higher reflectivity we observed suggests that the fluid
becomes conducting. 

%The result of shocked temperature was 0.78 $\pm$ 0.10 eV at 
%$U_{s{\rm H}} = 24.9 \pm 0.5$ km/s for experiment No. 32270. 
%The radiance and shock velocity were averaged over 0.8 ns during the
%shock is propagating in the hydrogen layer.
% The result of shocked temperature was 0.87, 0.76, 0.74, and 0.73 eV at 
% $U_{s{\rm H}} = 25.5$, 25.3, 24.7, and 24.4 km/s, respectively, for
% experiment No. 32270.  
% The radiance and shock velocity were averaged over every 0.2 ns during
% the shock is propagating in the hydrogen layer.
The Hugoniot temperature can be obtained using the reflectivity and SOP
spectral intensities.
An example of the intensity versus time data is shown in Fig. 5. 
The emission dropped dramatically when the shock front entered the
hydrogen layer from quartz.
Compared to the VISAR data, material boundary on SOP records is not
clear due to the poorer temporal resolution $\sim 0.2$ ns. 
%%%Thus the self-emission from the shock front is linearly fitted on
%%%longer time interval to derive an accurate Hugoniot temperature.
The self-emission profile of hydrogen just after the shock breakout is
then determined from extrapolating a linear fit to measurements back
to $t = 0$, in order to eliminate the contamination of the quartz
emission. 
The time average of the fitted intensity over the temporal resolution
is adopted for the derivation of the Hugoniot temperature of hydrogen.
The results of the hydrogen temperature are listed in Table II.
The size of the uncertainty in temperature is 6-19\%, which stems from
the system calibration, optical transmission, and measurements of
self-emission and reflectivity. 
%Theoretical models predicted that $T$ of shocked hydrogen is
%higher by a factor of $\sim$ 1.3 than that of deuterium at the same
%pressure \cite{kerley03}. 

Figure 6 depicts the Hugoniot temperature of liquid hydrogen as a
function of measured pressure.
Theoretical models predicted that Hugoniot temperatures of liquid
hydrogen are higher by a factor of $\sim 1.3$ than those of liquid
deuterium at the same pressure \cite{kerley03}. 
Deuterium temperature measured by previous shock experiments
\cite{holmes95,bailey08} are also shown in Fig. 6. 
%\cite{collins01prl}. 
Although the uncertainty in temperature measurements is still large,
the hydrogen temperature obtained in this work is higher than the
deuterium data.
This is another experimental evidence of the isotope difference
in the hydrogen Hugoniot.
At this temperature, molecular hydrogen begins to dissociate and the
fraction reaches to $\sim 14$\% (39\%) at $P = 25$ GPa (40 GPa)
according to Kerley's model \cite{kerley03}.  

\section{DISCUSSION AND CONCLUSIONS}

Drude-type models are often applied to parameterize the optical
properties of liquid metal \cite{hodgson72,celliers00}.
%\cite{hodgson72}.
Within the Drude description, the complex index of refraction is given
by $\hat{n}_s^2 = 1 - (\omega_p^2/\omega^2)(1 + i / \omega\tau_e)^{-1}$
where $\omega_p = (4 \pi n_e e^2 / m_e)^{1/2}$ is the plasma
frequency, $n_e$ is the carrier density, $e$ is the electron charge,
$m_e$ is the electron mass, and $\omega = 2 \pi c / \lambda$ is the
optical frequency.  
Here the electron relaxation time is assumed to be $\tau_e = R_0 / v_F$
where $R_0$ is the interparticle spacing and $v_F$ is the electron
Fermi velocity.
%This is appropriate for strongly scattering disordered systems
%\cite{lee84}. 
We adopt $R_0=0.126$ nm corresponding to a density $0.4$ g/cm$^3$. 
Then the Drude reflectivity of liquid metallic hydrogen can be derived
as a function of $n_e$, which is depicted in Fig. 7.
% In this figure, we set $R_0=0.126$ nm corresponding to a density $0.4$
% g/cm$^3$. 

High reflectivity is produced when the carrier density exceeds the
critical density $n_c$ defined by $\omega^2_p (n_c) = \omega^2$.
The critical density is $n_c = 3.9 \times 10^{21}$ cm$^{-3}$ at
$\lambda = 532$ nm. 
Taking $R = 23$\%, for example, the carrier density is given by $3.7
\times 10^{22}$ cm$^{-3}$, which is about 31\% of total hydrogen
number density $n_{\max}=1.2 \times 10^{23}$ cm$^{-3}$  and equivalent
with $\omega \tau_e = 0.37$. 
The Fermi energy estimated from this carrier density is 4.0 eV, which
is higher than the shock temperature.
This indicates that the observed highly reflective state would be
characteristic of a degenerate liquid metal.
% and not a high temperature plasma. 
Based on the Drude model, the reflectivity increases with the carrier
density and saturates at $R = 36$\%.
This implies that the maximum reflectivity of hydrogen may be slightly
lower than that of deuterium \cite{celliers00}, although it depends on
the assumed compression ratio.
% for the fully ionized case.
It will be interesting to confirm the saturated value of the
reflectivity by future experiments. 

The hydrogen Hugoniot obtained in this work is strongly dependent on
the quartz EOS. 
Recently the quartz Hugoniot was reexamined by using magnetically
driven flyer impact on the Z machine and a new fitting function was
derived \cite{knudson09}.
For comparison, the hydrogen Hugoniots calculated by using the Z-fit
and Kerley release ($U'_{p{\rm H}}$, $P'_{\rm H}$, and 
$\rho'_{\rm H}$) are listed in Table I.   
The stiffer Hugoniot of quartz reduces the initial pressure of the
release isentrope (see Fig. 2).  
At the initial shock state of quartz, the pressure and density
inferred from the Z-fit Hugoniot, $P_1 = 716$ GPa and $\rho_1 = 6.61$ 
g/cm$^3$, are lower than those of the OMEGA-fit case, $P_1 = 746$ GPa
and $\rho_1 = 7.05$ g/cm$^3$).  

In our analysis, the quartz release cannot be approximated by a
reflection of the Hugoniot in the $P$-$U_p$ plane, and thus the
off-Hugoniot EOS of Kerley's model determines the shape of the curve. 
The particle velocity along the release isentrope can be calculated by 
using the Riemann integral: 
\begin{equation}
U_p = U_{p1} - \int_{\rho_1}^{\rho} c_s \rho^{-1} d{\rho} \;,
\label{eqn:riemann}
\end{equation}
where $U_{p1}$ is the particle velocity at the initial shock state,
and the sound speed is obtained from the pressure derivative 
of the density at constant entropy $S$, $c_s^2 = (\partial  P /
\partial \rho)_S$. 
The second term of Eq. (\ref{eqn:riemann}) is larger for the Z-fit
case due to the lower density, and then the particle
velocity increases faster with the decrease of density. 
% Notice that the difference in the sound speed is small and $d{\rho}$
% is negative here.  
Therefore, two release curves shown in Fig. 2 are gradually
approaching at the lower pressure.  
The resulting differences in the Hugoniot pressure and density for
hydrogen are $\sim$ 1\% and $\sim$ 4\% compared to those derived by
the OMEGA-fit. 
% Although the shock Hugoniot is relatively well determined, EOS data
% along the release curves are still model dependent.
% For comparison, 
% a quartz release approximated by a reflection of
% the principal Hugoniot in the $P$-$U_p$ plane \cite{enig63} is shown
% in Fig. 2 and the corresponding  
% data, $U'_{p{\rm H}}$, $P'_{\rm H}$, and $\rho'_{\rm H}$, are listed
% in Table I.  
% For this case, systematic errors arising from uncertainties in the
% quartz Hugoniot are considered as well as random errors.
% At higher pressure, the mirror curve is higher than Kerley's model.
% But just above 50 GPa, they cross and the Kerley model gives a larger
% particle velocity.
% The Hugoniot pressure $P'_{\rm H}$ is lower by 2\% and the density
% $\rho'_{\rm H}$ lower by 6\% compared to the mirror approximation
% case. 
Apparently, improvement of the quartz EOS 
% including both the principal Hugoniot and off-Hugoniot 
is essential for the further development of
Hugoniot measurements using quartz standards \cite{boehly07}.  

In summary, we have obtained the principal Hugoniot data
$P$-$\rho$-$T$ for liquid hydrogen, not deuterium, in an unexplored
range of pressure up to 55 GPa.  
%Thanks to a quartz standard we measured shock temperature as well
%as the pressure and density simultaneously.
The results demonstrate that the hydrogen Hugoniot cannot be scaled by
density from the deuterium data in consequence of 
the initial density effects \cite{hicks09,militzer06,eggert08}.
%For astrophysical applications, experimental data of hydrogen, not
%deuterium, have quite important meanings.
As for the study of planetary interiors, the hydrogen EOS data at much
higher pressure are required since the transition to metallic hydrogen
is anticipated to be at $P \sim$ 200-400 GPa in Jupiter \cite{saumon04}. 
%and $T \sim $ 6000 - 9000 K 
%single shock compression is not sufficient for
%understanding of the interior of Jupiter.
% However, the increase of temperature must be suppressed significantly
% because the Hugoniot temperature at this pressure range is too high to
% reproduce Jupiter's conditions. 
However, the hydrogen temperature must be kept lower because the
Hugoniot temperature at this pressure range is too high to reproduce
Jupiter's conditions. 
Therefore off-Hugoniot measurements of hydrogen by means of reflection
shocks \cite{boehly04} and/or precompressed samples
\cite{eggert08,loubeyre04,kimura10} will be a quite important next step.
% \cite{loubeyre04,eggert08} will be a quite important next step.

% If in two-column mode, this environment will change to single-column
% format so that long equations can be displayed. Use
%\begin{widetext}
%\end{widetext}

% figures should be put into the text as floats.
% Use the graphics or graphicx packages (distributed with LaTeX2e)
% and the \includegraphics macro defined in those packages.
% See the LaTeX Graphics Companion by Michel Goosens, Sebastian Rahtz,
% entire page. There is no need to do explicit centering.

\begin{acknowledgments}
We thank G. Kerley, B. Holst, and M. Ross for providing us with their
EOS data and W. Nellis for useful discussions. 
We also thank the anonymous referees whose comments and suggestions
improved this paper.
We are deeply grateful to the GEKKO technical crew for their
exceptional support during these experiments and the Daico MFG Co. for
their outstanding work on quartz targets. 
%We also thank the GEKKO technical crew for their exceptional support
%during these experiments and the Daico MFG Co. for their outstanding
%work on quartz targets. 
This work was performed under the joint research project of the
Institute of Laser Engineering, Osaka University. 
This research was partially supported by grants for the Core-to-Core
Program from the Japan Society for the Promotion of Science,
the Global COE Program, ``Center for Electronic Devices
Innovation'', from the Ministry of Education, Culture, Sports, Science
and Technology of Japan, 
and the Core Research for Evolutional Science and Technology
(CREST) from the Japan Science and Technology Agency. 
\end{acknowledgments}

\clearpage 

\begin{figure}
\includegraphics[scale=0.85]{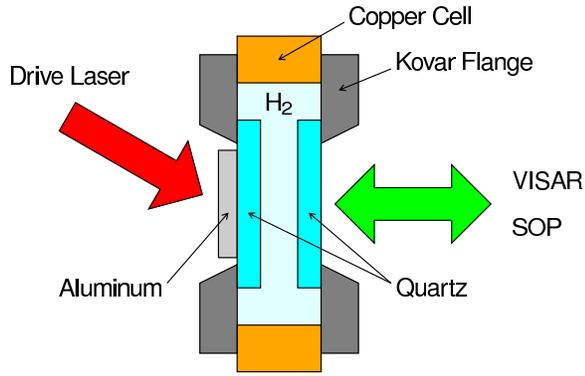}%
%\vspace{-5pt}
\caption{\label{f1}
(Color online) 
Sketch of the cryogenic hydrogen target used in the experiments. 
The drive laser irradiates the target from the left with an incident
angle of 30 degree, while the VISAR and SOP measure the shock velocity
and self-emission from the rear-side of the target.} 
\end{figure}

\begin{figure}
\includegraphics[scale=0.85]{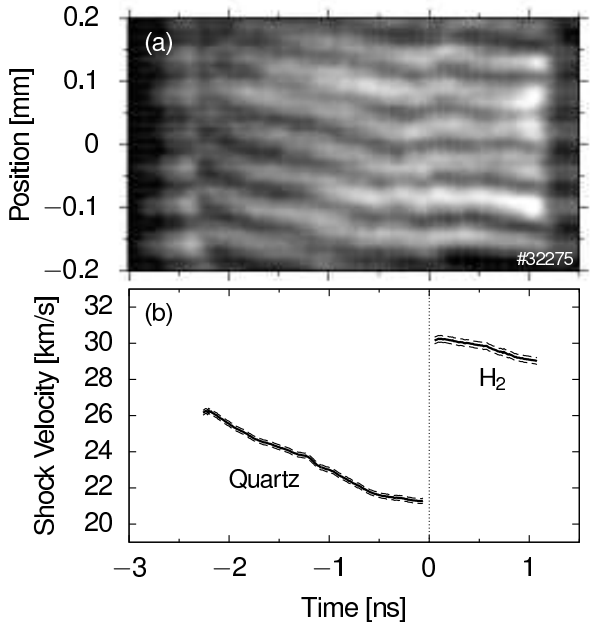}%
%\vspace{-5pt}
\caption{\label{f2}
(a) 
Sample VISAR trace showing the signal from the reflecting shock front
in the laser-side quartz and liquid hydrogen in experiment No. 32275. 
(b) 
Resulting velocity profile extracted from the VISAR fringe shift in
(a) with dashed lines representing measurement uncertainties.}  
\end{figure}

\begin{figure}
\includegraphics[scale=0.85]{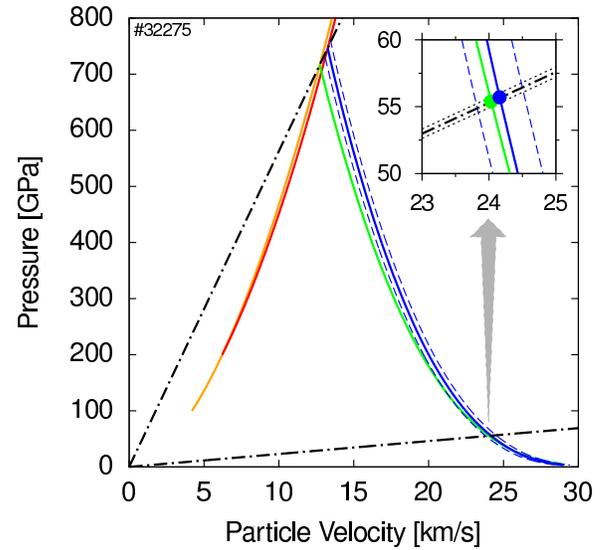}%
%\vspace{-5pt}
\caption{\label{f3}
(Color)
Impedance-matching in the $P$-$U_p$ plane illustrating measurement
error propagation. 
Black dot-dashed curves show the Rayleigh lines for quartz (upper) and
hydrogen (lower).
Red solid curve is the quartz Hugoniot derived by laser experiments
(Ref. 26).
%(Ref. \cite{hicks05}).
Blue solid curve is a release isentrope of quartz calculated from
Kerley's model (SESAME 7360) 
(Ref. 27), 
%(Ref. \cite{kerley99}), 
and blue dashed curves show the propagation of uncertainty in the
quartz shock velocity.  
The inset is a magnified view around the intersection of the Rayleigh
line of hydrogen with the release isentrope of quartz. 
Black dotted curves show the range of uncertainty in the Rayleigh
lines associated with the measurement error in hydrogen shock
velocity. 
For the purpose of comparison, the quartz Hugoniot based on Z
experiments 
(Ref. 35) 
%(Ref. \cite{knudson09}) 
and the corresponding Kerley
release are shown by orange and green solid curves.}  
\end{figure}

\begin{figure}
\includegraphics[scale=0.85]{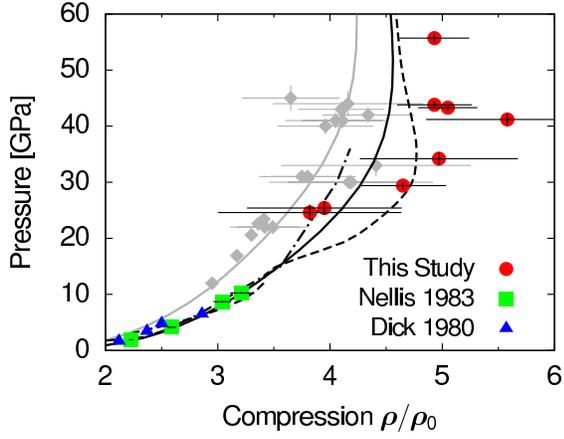}%
%\vspace{-5pt}
\caption{\label{f4}
(Color online)
Pressure versus compression for the principal Hugoniot of liquid
hydrogen. 
%The initial density is $\rho_0 = 0.0760$ g/cm$^{-3}$.
Data are from Dick and Kerley 
(Ref. 18) 
%(Ref. \cite{dick80}) 
(blue triangles), 
Nellis et al. 
(Ref. 19) 
%(Ref. \cite{nellis83}) 
(green squares), 
and this work (red circles).  
Error bars represent the quadrature sum of random and systematic
errors. 
%in the impedance-match analysis.
Also shown are theoretical predictions from the model EOS of Kerley
(Ref. 8) 
%(Ref. \cite{kerley03}) 
(solid line), 
quantum molecular dynamics simulations 
(Ref. 10)  
%(Ref. \cite{holst08})  
(dashed line), and the linear mixing model 
(Ref. 28) 
%(Ref. \cite{ross09}) 
(dot-dashed line).  
For reference, the Hugoniot data for liquid deuterium 
%(Refs. \cite{knudson04}, \cite{hicks09}, and \cite{holmes95}) 
(Refs. 14, 16, and 30) 
are depicted by gray diamonds, and gray curve is the Kerley model
for deuterium 
(Ref. 8).
%(Ref. \cite{kerley03}).
}  
\end{figure}

\begin{figure}
\includegraphics[scale=0.85]{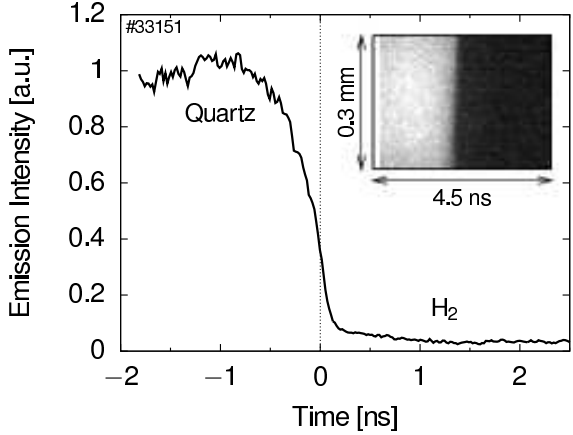}%
\caption{\label{f5}
Time profile of the emission intensity at 455 nm calculated from the
central region of the streaked shock emission data for experiment
No. 33151, which is shown in the inset image.
The thickness of the hydrogen layer was $\sim 250$ $\mu$m for this
shot, so that the transit time is much longer than 2.5 ns.}  
\end{figure}

\begin{figure}
\includegraphics[scale=0.85]{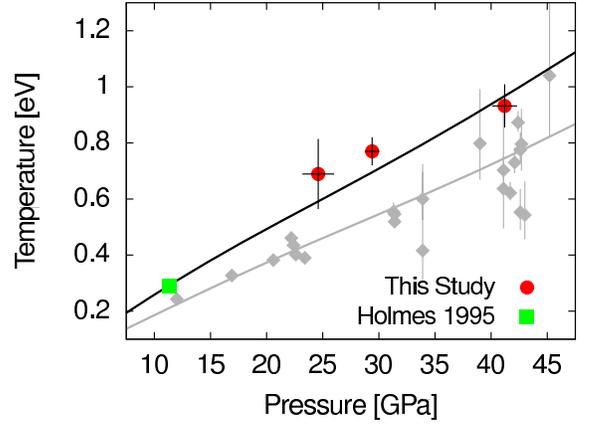}%
\caption{\label{f6}
(Color online)
Temperature versus pressure for the principal Hugoniot of liquid
hydrogen.
Experimental data are from Holmes et al. 
(Ref. 30)
%(Ref. \cite{holmes95})
(green squares) and this work (red circles).
For comparison, the Hugoniot temperature for liquid deuterium
(Refs. 30 and 32) 
%(Refs. \cite{holmes95} and \cite{bailey08}) 
are depicted by gray diamonds.
Also shown are theoretical predictions from Kerley's EOS model 
(Ref. 8) 
%(Ref. \cite{kerley03}) 
for hydrogen (solid line) and deuterium (gray line).} 
\end{figure}

\begin{figure}
\includegraphics[scale=0.85]{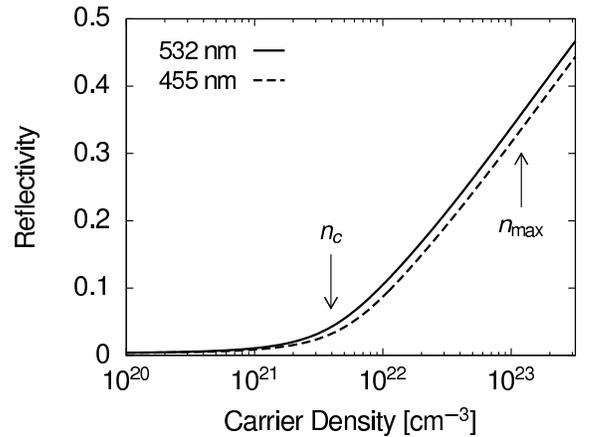}%
\caption{\label{f7}
Drude reflectivity plotted as a function of carrier density at the
VISAR wavelength 532 nm (solid line) and SOP wavelength 455 nm (dashed
line). 
The difference in the reflectivity between these two wavelengths is at
most 2\%. 
The critical density $n_c$ and maximum density $n_{\max}$ of electrons
are indicated by the arrows.}
\end{figure}

%\clearpage

\begin{turnpage}
\begin{table*}
%\begin{longtable}
\caption{
Hugoniot data for liquid hydrogen resulted from impedance-matching to
a quartz standard. 
$U_{s{\rm Q}}$ and $U_{s{\rm H}}$ are the shock velocities in quartz
and hydrogen with measurement errors. 
$U_{p{\rm H}}$, $P_{\rm H}$, and $\rho_{\rm H}$ are the particle
velocity, pressure, and density of shocked hydrogen showing both
random and systematic errors: (ran, sys). 
$U'_{p{\rm H}}$, $P'_{\rm H}$, and $\rho'_{\rm H}$ are reference data
derived by using the Z-fit quartz Hugoniot 
%(Ref. \cite{knudson09}) 
(Ref. 35) 
instead of the OMEGA laser fit 
(Ref. 26). 
%(Ref. \cite{hicks05}). 
Shot numbers with an asterisk indicate targets with a thinner aluminum
pusher (23 $\mu$m) and VISARs with lower sensitivities, 9.79 (13.51)
and 14.72 (20.31) km/s/fringe for quartz (hydrogen).} 
\begin{ruledtabular}
\begin{tabular}{lcccccccc}
Shot No. & $U_{s{\rm Q}}$ [km/s] &  $U_{s{\rm H}}$ [km/s] & 
$U_{p{\rm H}}$ [km/s] & $P_{\rm H}$ [GPa] &  $\rho_{\rm H}$ [g/cm$^3$]
& $U'_{p{\rm H}}$ [km/s] & $P'_{\rm H}$ [GPa] 
&  $\rho'_{\rm H}$ [g/cm$^3$] \\
32275 & $21.24 \pm 0.21$ & $30.31 \pm 0.20$ & $24.16 \pm (0.38, 0.01)$ &
$55.7 \pm (0.9, 0.0)$ &  $0.375 \pm (0.024, 0.001)$ & 
$24.03 \pm (0.39, 0.14)$ & $55.4 \pm (1.0, 0.3)$ & 
$ 0.367 \pm (0.023, 0.008)$ \\
33159 & $19.33 \pm 0.16$ & $26.89 \pm 0.30$ & $21.44 \pm (0.27, 0.04)$ &
$43.8 \pm (0.7, 0.1)$ &  $0.375 \pm (0.025, 0.003)$ & 
$21.29 \pm (0.29, 0.13)$ & $43.5 \pm (0.7, 0.3)$ & 
$ 0.365 \pm (0.025, 0.008)$ \\
33144 & $19.27 \pm 0.14$ & $26.64 \pm 0.20$ & $21.37 \pm (0.26, 0.05)$ &
$43.3 \pm (0.6, 0.1)$ &  $0.384 \pm (0.020, 0.003)$ & 
$21.21 \pm (0.27, 0.13)$ & $42.9 \pm (0.6, 0.3)$ & 
$ 0.373 \pm (0.020, 0.009)$ \\
32270 & $19.04 \pm 0.34$ & $25.71 \pm 0.25$ & $21.10 \pm (0.54, 0.06)$ &
$41.2 \pm (1.1, 0.1)$ & $0.424 \pm (0.054, 0.005)$ & 
$20.92 \pm (0.56, 0.13)$ & $40.9 \pm (1.2, 0.3)$ & 
$ 0.408 \pm (0.054, 0.011)$ \\
31912$^{*}$ & $17.54 \pm 0.28$ & $23.72 \pm 0.45$ & $18.95 \pm (0.49, 0.07)$ & 
$34.2 \pm (1.1, 0.1)$ & $0.378 \pm (0.053, 0.006)$ & 
$18.77 \pm (0.44, 0.20)$ & $33.8 \pm (1.0, 0.4)$ & 
$ 0.364 \pm (0.045, 0.014)$ \\
33151 & $16.58 \pm 0.17$ & $22.21 \pm 0.27$ & $17.43 \pm (0.32, 0.11)$ &
$29.4 \pm (0.6, 0.2)$ & $0.353 \pm (0.028, 0.008)$ & 
$17.50 \pm (0.28, 0.22)$ & $29.5 \pm (0.6, 0.4)$ & 
$ 0.358 \pm (0.027, 0.016)$ \\
31917$^{*}$ & $15.49 \pm 0.44$ & $21.14 \pm 0.54$ & $15.78 \pm (0.68, 0.14)$ &
$25.4 \pm (1.2, 0.2)$ & $0.300 \pm (0.052, 0.008)$ & 
$15.97 \pm (0.67, 0.17)$ & $25.7 \pm (1.2, 0.3)$ & 
$ 0.311 \pm (0.054, 0.010)$ \\
31922$^{*}$ & $15.26 \pm 0.48$ & $20.93 \pm 0.71$ & $15.45 \pm (0.74, 0.14)$ &
$24.6 \pm (1.4, 0.2)$ & $0.290 \pm (0.062, 0.008)$ & 
$15.64 \pm (0.74, 0.14)$ & $24.9 \pm (1.4, 0.2)$ & 
$ 0.301 \pm (0.066, 0.008)$ \\
\end{tabular}
\end{ruledtabular}
%\endnote{For
%  shot number 31912 and 31917, the lower velocity sensitivities were
%  used: This cause relatively larger measurement errors in shock
%  velocities.}.  
%\end{longtable}
\end{table*}
\end{turnpage}

\begin{table}
\caption{
Reflectivity $R_{\rm H}$ and Hugoniot temperature $T_{\rm H}$ for
liquid hydrogen. 
The reflectivity of shocked hydrogen was measured at the
wavelength 532 nm using VISAR signal and aluminum known reflectivity.  
The Hugoniot pressure and density for each shot are listed in Table I.}   
\begin{ruledtabular}
\begin{tabular}{lccccccc}
Shot No. & $R_{\rm H}$ [\%] & $T_{\rm H}$ [eV] \\
32270 & $22.7 \pm 2.8$ & $0.93 \pm 0.08$ \\
33151 & $14.1 \pm 1.9$ & $0.77 \pm 0.05$ \\
31922 & $8.3 \pm 1.8$ & $0.69 \pm 0.13$ \\
\end{tabular}
\end{ruledtabular}
\end{table}

% Surround figure environment with turnpage environment for landscape
% figure
% \begin{figure}
% \end{figure}

% tables should appear as floats within the text
% The ruledtabular enviroment adds doubled rules to table and sets a
% reasonable default table settings.
% Use the table* environment to get a full-width table in two-column
% Add \usepackage{longtable} and the longtable (or longtable*}
% environment for nicely formatted long tables. Or use the the [H]
% placement option to break a long table (with less control than 
% \begin{table}%[H] add [H] placement to break table across pages
% \begin{ruledtabular}
% Lines of table here ending with \\
% \end{tabular}
% \end{ruledtabular}
% \end{table}

% Surround table environment with turnpage environment for landscape
% table
% \begin{table}
% \begin{ruledtabular}
% \end{tabular}
% \end{ruledtabular}
% \end{table}

%\appendix
%\section{}

\clearpage

%\bibliography{material}

%\clearpage

\end{document}